\documentclass[showpacs,aps,graphicx,twocolumn]{revtex4}
\usepackage{amsmath}
\usepackage{graphicx}
\usepackage{txfonts}

\begin{document}


\title{Shortcuts to adiabatic holonomic quantum computation in decoherence-free subspace
with transitionless quantum driving algorithm\footnote{Published in
New J. Phys. \textbf{18}, 023001 (2016)} }

\author{Xue-Ke Song, Hao Zhang, Qing Ai, Jing Qiu, and Fu-Guo Deng\footnote{Corresponding author: fgdeng@bnu.edu.cn}}

\address{Department of Physics, Applied Optics Beijing Area Major Laboratory,
Beijing Normal University, Beijing 100875, China}

\date{\today }

\begin{abstract}
By using transitionless quantum driving algorithm (TQDA), we present
an efficient scheme for the shortcuts to the holonomic quantum
computation (HQC). It works in decoherence-free subspace (DFS) and
the adiabatic process can be speeded up in the shortest possible
time. More interestingly, we give a physical implementation for our
shortcuts to HQC with nitrogen-vacancy centers in diamonds
dispersively coupled to a whispering-gallery mode microsphere
cavity. It can be efficiently realized by controlling appropriately
the frequencies of the external laser pulses. Also, our scheme has
good scalability with more qubits. Different from  previous works,
we first use TQDA to realize a universal HQC in DFS, including not
only two noncommuting accelerated single-qubit holonomic gates but
also a accelerated two-qubit holonomic controlled-phase gate, which
provides the necessary shortcuts for the complete set of gates
required for universal quantum computation. Moreover, our
experimentally realizable shortcuts require  only two-body
interactions, not four-body ones, and they work  in the dispersive
regime, which relax greatly the difficulty of their physical
implementation in experiment. Our numerical calculations show that
the present scheme is robust against decoherence with current
experimental parameters.
\end{abstract}


\maketitle


\section{Introduction}
\label{sec1}

Quantum computation (QC), which permits unitary operations on
qubits, has attracted considerable attention in recent years
\cite{computation}. Many interesting theoretical schemes have been
proposed for universal quantum logic gates in various quantum
systems, such as trapped ions \cite{ion}, atom-cavity systems
\cite{atom}, photons \cite{KLM,hyperCNOT2}, quantum dots \cite{QD1},
circuit quantum electrodynamics \cite{circuitQED1}, and so on. In
experiment, there are  stochastic control errors during the gate
operation and the collective noise caused by the interaction between
a quantum system and its ambient environment. To suppress the
former, Zanardi and Rasetti \cite{HQC1} introduced the holonomic
quantum computation (HQC) which is based on the adiabatic
non-abelian geometric phases (holonomies) in 1999.  The advantage of
HQC is that it depends only on the global geometric properties of
the evolution in parameter space, but resilience of the local noises
and fluctuations \cite{HQC1,HQC2}. In 2001, Duan \emph{et al.}
\cite{Duan}  proposed an interesting scheme for adiabatic geometric
QC in trapped ions. Subsequently, much effort was made on
nonadiabatic geometric QC
\cite{NHQC1,NHQC2,NHQC3,NHQC4,NHQC5,NHQC6,NHQC7,NHQC8,NHQC9,NHQC10,NHQC11}
and unconventional geometric QC \cite{UHQC1,UHQC2,UHQC3}. These
holonomic quantum gates are more robust than the conventional ones.
Interestingly, the nonadiabatic geometric QC were demonstrated in
several physical systems by some groups. For example, in 2013, Feng,
Xu, and Long \cite{NHQCLong} experimentally realized the
nonadiabatic HQC in a liquid NMR quantum information processor for
the first time, including one-qubit holonomic gates and the
two-qubit holonomic controlled-not gate. Meanwhile, Abdumalikov
\emph{et al.} \cite{NHQCSconducting} realized firstly the
nonadiabatic holonomic single-qubit  operations on a three-level
transmon qubit. In 2014, two groups \cite{NHQCNV,NHQCNVDuan}
demonstrated the nonadiabatic holonomic quantum gates in diamond
nitrogen-vacancy (NV) centers.

The decoherence-free subspace (DFS) \cite{DFS1,DFS2,DFS3} of a
quantum system can protect the fragile quantum information against
collective noises as the system undergoes a unitary evolution in its
DFS. It has been demonstrated that DFS can be implemented
experimentally with different physical systems
\cite{DFSp1,DFSp2,DFSp3}. In 2005, Wu \emph{et al.} \cite{LianaoWu}
presented a theoretic scheme by combining the HQC and DFS to perform
universal QC. By making the dark states of the Hamiltonian of a
quantum system adiabatically evolve along a closed cyclic loop, one
can acquire a Berry phase or quantum holonomy. In 2006, Zhang
\emph{et al.} \cite{XinDingZhang} and Cen \emph{et al.}
\cite{LiXiangCen} gave two schemes for HQC with DFS in trapped ions.
In 2009, Oreshkov \emph{et al.} \cite{HQCED1} introduced a scheme
for fault-tolerant HQC on stabilizer codes. The adiabatic evolution
for HQC requires a long run time. To eliminate this dilemma, Berry
\cite{Berry} came up with a transitionless quantum driving algorithm
(TQDA), which is also outlined in slightly different manner by
Demirplak and Rice \cite{Demirplak1,Demirplak2}, to speed up the
adiabatic quantum gates when the eigenstates of a time-dependent
Hamiltonian are non-degenerate in 2009. Later, this transitionless
algorithm has been gained widespread attention in both theory and
experiment
\cite{TQDAthree,TQDAnonlinear,TQDAtrack,TQDAlevel,TQDAexperiment,TQDANV}.
In 2010, Chen \emph{et al.} \cite{TQDAthree} used the TQDA to speed
up adiabatic passage techniques in two-level and three-level atoms
extending to the short-time domain their robustness with respect to
parameter variations. In 2012, Bason \emph{et al.}
\cite{TQDAexperiment} experimentally implemented the optimal
high-fidelity transitionless superadiabatic protocol on
Bose-Einstein condensates in optical lattices. In 2013, Zhang
\emph{et al.} \cite{TQDANV} implemented the acceleration of quantum
adiabatic passages on the electron spin of a single NV center in
diamond. As for the degenerate case, Zhang \emph{et al.}
\cite{JiangZhang} generalized TQDA to show the adiabatic shortcuts
to holonomic quantum gates without DFS. In 2015, Pyshkin \emph{et
al.} \cite{Pyshkin} showed that the conventional HQC can be
accelerated by using external control fields.

Recently, the diamond NV center coupled to a quantized
whispering-gallery mode (WGM) of a fused-silica high-Q microcavity
has been extensively investigated in quantum information. On one
hand, an NV center in a diamond has long electron-spin coherence
time even at room temperature \cite{NV1}, and it is  easy to
manipulate, initialize, and readout the quantum state on the NV
center via the external laser and microwave field
\cite{NHQCNV,NHQCNVDuan}. On the other hand, a microcavity can
attain a ultrahigh Q factor ($\,>\!\!\!10^{8}$ even up to $10^{10}$)
with a very small volume \cite{Microsphere1,Microsphere2}. Taking
advantage of the exceptional spin features of NV centers and the
ultrahigh-Q factor of microsphere cavity, Park \emph{et al.}
\cite{coupling1} observed the normal mode splitting in this cavity
QED system in 2006. Afterward, some interesting schemes for
high-fidelity entanglement generation between separate NV centers
and other quantum information tasks have been proposed
\cite{generation1,weiNVgate,hyperCNOT3,hentangle}. In 2010, Yang
\emph{et al.} \cite{generation1} proposed a scheme for generating
the W state and Bell state in this nanocrystal-microsphere system.
In 2015, Ren \emph{et al.} \cite{hyperCNOT3} presented the dipole
induced transparency of an NV center embedded in a photonic crystal
cavity coupled to two waveguides and designed two universal
hyperparallel hybrid photonic quantum logic gates. Liu and Zhang
\cite{hentangle} proposed two efficient schemes for the
deterministic generation and the complete nondestructive analysis of
hyperentangled Bell states, assisted by the NV centers coupled to
microtoroidal resonators.

In this paper, we propose an efficient scheme to speed up the
adiabatic holonomic quantum gates in DFS by using TQDA. This
proposal takes advantage of the fault tolerance of HQC and coherence
preserving virtues of DFS to protect quantum information from local
fluctuations and collective noises. The TQDA makes the adiabatic
holonomic quantum process be accelerated in the shortest possible
time. In addition, we present a feasible physical implementation of
this protocol with diamond NV centers dispersively coupled to a
quantized WGM of a microsphere cavity. We can achieve the shortcuts
to adiabatic HQC in DFS by tuning the frequencies of the external
laser field, which simplifies the operation procedure largely. Our
scheme is scalable as it can be straightforwardly applied to HQC
with multiple qubits. Different from  previous works, we use TQDA to
realize a universal HQC in DFS, including both two noncommuting
accelerated single-qubit holonomic gates and a accelerated two-qubit
holonomic controlled-phase (CP) gate. This provides an efficient
route for shortcuts to adiabatic HQC in DFS. Moreover, the present
proposal requires only two-body interaction, not four-body ones,
which largely reduces the experimental challenge. With a virtual
photon process, the cavity decay is greatly suppressed. Our
numerical calculations show that this scheme can reach a high
fidelity with current experiment parameters, and it exhibits the
robustness of the HQC.

\section{Basic theories}
\label{sec2}

Let us give a brief review of TQDA for a general quantum system with
an arbitrary time-dependent Hamiltonian $H_{0}(t)$. If the initial
state is in one of the eigenstates of the Hamiltonian $H_{0}(t)$,
the quantum adiabatic theorem guarantees that the system remains
approximately in this eigenstate when the time evolution is
sufficiently slow. Due to the long runtime required for adiabatic
evolution, it will bring in the extra loss of coherence and
spontaneous emission of the quantum system. In 2009, Berry
\cite{Berry} introduced an optimized quantum algorithm, i.e., TQDA,
to speed up the adiabatic process. Actually, the main idea of TQDA
is that if the adiabatic approximation for the evolution operator of
a given quantum system is specified, one can find another
Hamiltonian $H(t)$ which can generate the equivalent unitary
transformation in a shortest possible time. In the TQDA, the
Hamiltonian $H(t)$ drives the evolving states following the selected
instantaneous adiabatic eigenstates of $H_{0}(t)$ exactly without
undergoing transitions, while there is no limitation to the
adiabatic theorem. The Hamiltonian $H(t)$ can be divided into two
parts: one is the fundamental Hamiltonian $H_{0}(t)$ for adiabatic
evolution, and the other is the additional Hamiltonian $H_{1}(t)$
which can suppress the transitions of the system due to the rapid
evolution. In theory, the TQDA ensures that the state of the quantum
system remains in the eigenstate of the Hamiltonian $H_{0}(t)$
invariable for all time. This reveals that the adiabatic evolution
can be accelerated close to the quantum speed limit by using the
TQDA \cite{{TQDAexperiment}}.

More specifically, considering the fundamental Hamiltonian
$H_{0}(t)$ of the quantum system has the non-degenerate
instantaneous eigenstates $|n(t)\rangle$ with corresponding
eigenvalues $E_{n}(t)$, in the adiabatic approximation, one could
write the state evolution of the system by
\begin{eqnarray}    
|\Psi_{n}(t)\rangle=exp\left\{-i\!\!\int_{0}^{t}\!\!dt^{'}\!E_{n}(t^{'}) 
-\!\!\int_{0}^{t}\!\!dt^{'}\langle n(t^{'})|\dot{n}(t^{'})\rangle\right\}|n(t)\rangle.
\end{eqnarray}
By employing the reverse engineering approach described in Ref.
\cite{Berry}, the driving Hamiltonian $H(t)$, with $\hbar=1$, takes
the form of
\begin{eqnarray}    
H(t)=\!\!\sum_{n}\!E_{n}|n\rangle\langle
n|\!+\!i\sum_{n}(|\dot{n}\rangle \langle n|\!-\!\langle
n|\dot{n}\rangle|n\rangle\langle
n|)=\!H_{0}(t)\!+\!H_{1}(t),\nonumber\\
\!\!\!\!\!\!\!\!\!\!\!\!\!\!\!\!\!\!\!\!\!\!\!\!\!\!\!\!\!\!\!\!
\end{eqnarray}
where all kets are time-dependent and
$H_{0}(t)=\sum_{n}E_{n}|n\rangle\langle n|$.

On the other hand, if there exists degeneracy in the spectrum of the
Hamiltonian, the situation becomes more complex and troublesome. To
get rid of this dilemma, Zhang \emph{et al. } \cite{JiangZhang}
generalized the non-degenerate TQDA to the degenerate case, which
can acquire non-Abelian geometric phases, i.e., quantum holonomies,
after a cyclic evolution. Likewise, the transitionless driving
Hamiltonian to achieve adiabatic shortcuts for the degenerate case
is given by
\begin{eqnarray}    
H^{\,'}(t)\!&=&\!\sum_{n,k}E_{n}\,|\varphi_{k}^{n}\rangle\langle\varphi_{k}^{n}|
+\sum_{n}\left(i\,|\dot{\varphi}_{k}^{n}\rangle\langle\varphi_{k}^{n}|
-A_{kl}^{n}\,|\varphi_{k}^{n}\rangle\langle\varphi_{l}^{n}|\right)\nonumber\\
\!&=&\!H_{0}^{\,'}(t)+H_{1}^{\,'}(t),\;\;\;\;\;\;\;\;\;\label{shortcut}
\end{eqnarray}
in which
$H_{0}^{\,'}(t)=\sum_{n,k}E_{n}\,|\varphi_{k}^{n}\rangle\langle\varphi_{k}^{n}|$
, $|\varphi_{k}^{n}\rangle$ ($k=1,2,...,m_{n}$) are a set of
degenerate eigenstates with the corresponding eigenvalues $E_{n}(t)$
of the Hamiltonian $H_{0}^{\,'}(t)$, and
$A_{kl}^{n}=i\,\langle\varphi_{k}^{n}|\dot{\varphi}_{l}^{n}\rangle$
represents the matrix-valued connection, also known as the holonomy
matrix.

\begin{figure}[!h]
\begin{center}
\includegraphics[width=8 cm,angle=0]{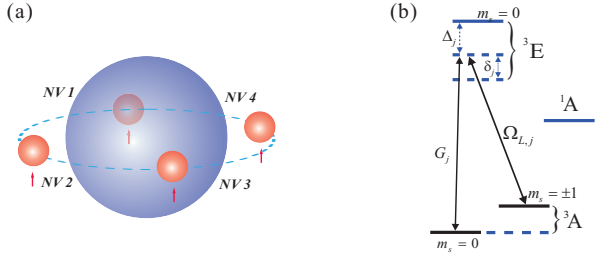}
\caption{(a) Schematic diagram for $N$ identical NV centers locating
around the equator of a fused-silica microsphere cavity. (b) The
energy-level configuration for an NV center, where $\Delta_{j}$ and
$\delta_{j}$ are the detunings, $G_{j}$ and $\Omega_{L,j}$ are the
coupling strength between an NV center and a quantized WGM of the
microsphere cavity and that between an NV center and the external
laser field, respectively. Here, the states
$|\,^{3}A,m_{s}=0\rangle$, $|\,^{3}A,m_{s}=-1\rangle$ and
$|\,^{3}E,m_{s}=0\rangle$ are encoded as the qubit states
$|0\rangle$, $|1\rangle$, and $|e\rangle$, respectively.}
\label{fig1}
\end{center}
\end{figure}

\section{Effective Hamiltonian based on NV centers interacting with microsphere resonator}
\label{sec3}

Our system is composed of \emph{N} identical NV centers in \emph{N}
separate diamond nanocrystals which are dispersively coupled to a
quantized WGM at the equator of single fused-silica microsphere
cavity, respectively, shown in Fig. \ref{fig1}(a). An NV center
consists of a substitutional nitrogen atom and an adjacent vacancy
in diamond lattice, and it can be easily manipulated by optical and
microwave field. By imposing laser pulses on the arbitrary NV center
interacting with the WGM, the NV center can be modeled as a
$\Lambda$-type three-level structure, as shown in Fig.
\ref{fig1}(b), where the states $|\,^{3}A,m_{s}=0\rangle$ and
$|\,^{3}A,m_{s}=-1\rangle$ are labeled by the qubit states
$|0\rangle$ and $|1\rangle$, respectively. $|\,^{3}E,m_{s}=0\rangle$
serves as the excited state $|e\rangle$. In our scheme, the
transition $|0\rangle\leftrightarrow|e\rangle$ with the frequency
$\omega_{e0}$ is far-off resonant with the WGM whose frequency is
$\omega_{c}$, and $|1\rangle\leftrightarrow|e\rangle$ with the
frequency $\omega_{e1}$ is driven by a largely detuned classical
laser field with the frequency $\omega_{L}$ and the polarization
$\sigma^{+}$ \cite{polarization}. Assuming both the coupling
strengths $G_{j}$ and $\Omega_{L,j}$ are sufficiently smaller than
the detuning $\Delta_{j}$, the state $|e\rangle$ can be
adiabatically eliminated. The NV centers are fixed and separated by
distance much larger than the wavelength of the WGM, so that there
are no the direct coupling among NV centers, and they can interact
with laser beams individually. Under the rotating wave
approximation, the interaction Hamiltonian, in the interaction
picture, can be expressed as
\begin{eqnarray}    
H_{int}=\sum_{j=1}^{N}g_{j}\,a\sigma_{j}^{+}\,e^{-\,i\,(\delta_{j}\,t\,-\,\phi_{j})}+H.c.,
\label{effective}
\end{eqnarray}
where $a^{+}$ ($a$) is the creation (annihilation) operator for the
WGM, $\phi_{j}$ is initial phase of the laser field imposed on the
$j$-th NV center, $\sigma_{j}^{+}=|1\rangle_j\langle 0|$,
$\sigma_{j}^{-}=|0\rangle_j\langle 1|$, and the coupling strength
$g_{j}=G_{\!j}\,\Omega_{L,j}\left(\frac{1}{\Delta_{j}+\delta_{j}}+\frac{1}{\Delta_{j}}\right)$
with $\Delta_{j}=\omega_{e0,j}-\omega_{c}$,
$\delta_{j}=\omega_{c}-\omega_{10,j}-\omega_{L,j}$, and
$\omega_{10,j}=\omega_{e0,j}-\omega_{e1,j}$.

To take $\delta_{j} \gg g_{j}$ into account, i.e., in the dispersive
regime, the direct energy exchange between NV centers and WGM is
negligible. Using the unitary transformation
$U=exp\left[\frac{g}{\delta}(a^{+}\sigma^{-}-a\sigma^{+})\right]$ to
eliminate the direct NV-center-WGM coupling, one can obtain the
effective Hamiltonian for the system composed of the NV centers as
follows:
\begin{eqnarray}    
H_{eff}=\sum_{j=1}^{N}\frac{g_{j}^2}{\delta_{j}}aa^{+}|1\rangle_j\langle
1| +\!\!\sum_{j,\,k,j\neq
k}^{N}\!\!\lambda_{j,\,k}\left(e^{i\,\phi_{j\,k}}\sigma_{j}^{-}\sigma_{k}^{+}+H.c.\right),\;\;
\end{eqnarray}
where
$\lambda_{j,\,k}=\frac{g_{j}\,g_{k}}{2}\left(\frac{1}{\delta_{j}}+\frac{1}{\delta_{k}}\right)$
and $\phi_{j\,k}=\phi_{j}-\phi_{k}$. Here we assume the WGM filed is
initially in the vacuum state. The first term corresponds to the
Stark shift term, which can be compensated by applying additional
lasers with appropriate frequencies
\cite{additionallaser1,additionallaser2}. For simplicity, hereafter,
we assume that the coupling strengths $g_{j}$ ($j=1,2,\cdots$) are
identical for all the NV centers, that is, $g_{j}=g_{k}=g$. The
effective Hamiltonian can be further simplified as
\begin{eqnarray}    
H\,'_{eff}=\sum_{j,\,k,j\neq
k}^{N}\lambda\,'_{j,\,k}\left(e^{\,i\,\phi_{j\,k}}\sigma_{j}^{-}\sigma_{k}^{+}+H.c.\right),
\end{eqnarray}
where
$\lambda\,'_{j,\,k}=\frac{g^2}{2}\left(\frac{1}{\delta_{j}}+\frac{1}{\delta_{k}}\right)$,
which serves as the effective Rabi frequency for the energy
conservation transition between the $j$-th and $k$-th NV centers. It
is indicated that the Rabi frequency $\lambda\,'_{j,\,k}$ is
inversely proportional to the detuning $\delta_{j}$ ($\delta_{k}$),
which relies largely on the difference between the frequencies of
cavity field $\omega_{c}$ and the external laser field
$\omega_{L,j}$ ($\omega_{L,\,k}$). With the Hamiltonian
$H\,'_{eff}$, applying different initial conditions, one can achieve
efficiently the shortcuts to the adiabatic single-logic-qubit gates
and the two-logic-qubit gate in DFS on this NV-center system, and
the detailed physical implementation about it will be presented in
next Section.

\section{Shortcuts to adiabatic single-qubit holonomic gates in DFS with NV centers system} \label{sec4}

\subsection{Shortcuts to adiabatic single-qubit bit-phase gate in DFS}
\label{sec41}

Considering a four-NV-center system which is coupled to a
microsphere resonator  in a symmetric way and undergoes a dephasing
process, described by the interaction Hamiltonian
$H_{I}=\sum_{i=1}^{4}\sigma_{z}^{i}\otimes B$, where $B$ is an
arbitrary environment operator. The DFS against the collective
dephasing noise can be expressed as
$C_{1}:=span\{|0001\rangle,|0010\rangle,|0100\rangle,|1000\rangle\}$,
in which $|0\rangle_{L}=|0001\rangle$ and
$|1\rangle_{L}=|0010\rangle$ donate the computational basis, and the
remaining states $|a_1\rangle=|1000\rangle$ and
$|a_2\rangle=|0100\rangle$ are employed as the ancillary states. For
accomplishing the singe-qubit bit-phase gate in this logical DFS, we
can utilize the target Hamiltonian
\begin{eqnarray}    
H^{y}_{0}(t)=\lambda'_{1,4}\,|a_1\rangle_{L}\langle 0| + \lambda'_{1,3}\,|a_1\rangle_{L}\langle 1| + \lambda'_{1,2}\,|a_1\rangle\langle a_2|+H.c.,
\end{eqnarray}
where $\lambda'_{j,k}$ is the effective coupling strength  between
the $j$-th and $k$-th NV centers for this four-NV-center system,
$\lambda'_{1,4}=\lambda'\sin\theta\cos\varphi$,
$\lambda'_{1,3}=\lambda'\sin\theta\sin\varphi$,
$\lambda'_{1,2}=\lambda'\cos\theta$ with $\lambda'=\sqrt{|\lambda'
_{1,2}|^2+|\lambda' _{1,3}|^2+|\lambda' _{1,4}|^2}$, and $\theta$
and $\varphi$ are the time-dependent tunable parameters with
$\theta\in[0,\pi]$ and $\varphi\in[0,2\pi]$. The dark states of the
Hamiltonian $H^{y}_{0}(t)$ are
$|D\,'_{0}(t)\rangle=\cos\theta\cos\varphi|0\rangle_{L}+\cos\theta\sin\varphi|1\rangle_{L}-\sin\theta|a_2\rangle$
and
$|D\,'_{1}(t)\rangle=-\sin\varphi|0\rangle_{L}+\cos\varphi|1\rangle_{L}$,
respectively. Under the adiabatic cyclic evolution of the dark
states, one gets the required single-qubit holonomic bit-phase gate
$U_y=e^{i\,\beta_{2}\,\sigma^{y}}$, where
$\sigma^{y}=i\,(|0\rangle_{L}\langle 1|-|1\rangle_{L}\langle 0|)$
and $\beta_{2}$ is the Berry phase factor. As shown in
Eq.~(\ref{shortcut}), it is known that with the purpose of achieving
shortcuts to the adiabatic gate $U_y$, one needs an additional
Hamiltonian $H^{y}_{1}(t)$ that can block the transition of quantum
states caused by the rapid evolution of the system. In the ordered
orthogonal basis
$\{|a_1\rangle,|0\rangle_{L},|1\rangle_{L},|a_2\rangle\}$, the
additional Hamiltonian for speeding up the adiabatic single-qubit
holonomic bit-phase gate $U_y$ reads
\begin{eqnarray}    
H^{y}_{1}(t)\!&=&\! i\cos\theta\,\dot{\varphi} 
\left(
                                         \begin{array}{cccc}
                                           0 & 0 & 0 & 0 \\
                                           0 & 0 & \cos\theta & -\sin\theta\sin\varphi \\
                                           0 & -\cos\theta & 0 & \sin\theta\cos\varphi \\
                                           0 & \sin\theta\sin\varphi & -\sin\theta\cos\varphi & 0 \\
                                         \end{array}
                                       \right)\nonumber\\
            &&+\;i\,\left(
                 \begin{array}{cccc}
                   0 & 0 & 0 & 0 \\
                   0 & 0 & -\dot{\varphi} & \cos\varphi\,\dot{\theta} \\
                   0 & \dot{\varphi} & 0 & \sin\varphi\,\dot{\theta} \\
                   0 & -\cos\varphi\,\dot{\theta} & -\sin\varphi\,\dot{\theta} & 0 \\
                 \end{array}
               \right).
\end{eqnarray}

Now, we focus on how to realize the shortcuts to the adiabatic
single-logic-qubit bit-phase gate $U_y$ on the four-NV-center system
by making use of the additional Hamiltonian. In this case, we choose
the initial phase difference between the external classical laser
pulses $\phi_{j\,k}=0$. On the first step, setting $\varphi=0$,
while increasing $\theta$ from $0$ to $\pi/2$, the corresponding
Hamiltonian of the system in the DFS $C_{1}$  is
$H^{y_1}=\lambda'\sin\theta\,|a_1\rangle_{L}\langle
0|+\lambda'\cos\theta\,|a_1\rangle \langle
a_2|+i\,\dot{\theta}\,|0\rangle_{L}\langle a_2|+H.c.$. This
Hamiltonian is under a $\Delta$-style structure. For achieving this
goal, we can tune the effective Rabi frequencies between the
physical qubits as $\lambda'_{1,4}=\lambda'\sin\theta$,
$\lambda'_{1,2}=\lambda'\cos\theta$, and
$\lambda'_{2,4}=\dot{\theta}$, and other control parameters are
zero. In other words, the first step can be effectively completed by
the three laser fields with different frequencies being applied to
the $1$-st, $2$-nd, and $4$-th NV centers, while there is no any
operation on $3$-th NV center when the cavity frequency is constant.
Second, keeping $\theta$ invariant but changing $\varphi$ from $0$
to a certain value $\varphi_{c}$, the required Hamiltonian takes the
form of $H^{y_2}=\lambda'\cos\varphi\,|a_1\rangle_{L}\langle
0|+\lambda'\sin\varphi\,|a_1\rangle_{L}\langle
1|-i\,\dot{\varphi}\,|0\rangle_{LL}\langle 1|+H.c.$. Adjusting the
effective Rabi frequencies $\lambda_{1,4}=\lambda'\cos\varphi$,
$\lambda_{1,3}=\lambda'\sin\varphi$, and
$\lambda_{3,4}=\dot{\varphi}$, one can obtain the required
Hamiltonian. It is worth emphasizing that the minimal resources with
three different frequencies of the external laser pulse can achieve
the second step. Finally, we keep $\varphi$ unchanged while decrease
$\theta$ to $0$. The control Hamiltonian for this case reads
$H^{y_3}=\lambda'\sin\theta\cos\varphi\,|a_1\rangle_{L}\langle
0|+\lambda'\sin\theta\sin\varphi\,|a_1\rangle_{L}\langle
1|+\lambda'\cos\theta\,|a_1\rangle\langle a_2|+
i\,\dot{\theta}\cos\varphi\,|0\rangle_{L}\langle
a_2|+i\,\dot{\theta}\sin\varphi\,|1\rangle_{L}\langle a_2|+H.c$.,
and then the system forms a cyclic evolution after tuning $\varphi$
to $0$. Different from the former two steps, in order to realize the
last step, all of the four NV centers should be imposed on the
external laser pulse with different frequencies to obtain the
different effective Rabi frequencies $\lambda'_{j,k}$ ($j,
k=1,2,3,4$, and $j\neq k$). The details for the parameters chosen in
each step for speeding up the adiabatic $U_y$ gate are shown in
Table \ref{table1}. Up to now, we have implemented the shortcuts to
the single-logic-qubit holonomic bit-phase gate on the
four-NV-center system in DFS.

\begin{table}[htb]
\centering \caption{Scheme for a three-step approach to realize the
shortcuts to the adiabatic holonomic single-qubit bit-phase gate.}
\begin{tabular}{cccc}
\hline\hline
  Step                                &        $\;\;\;\;\;\;$ $\theta$                     & $\;\;\;\;\;\;$ $\varphi$        &   the required Hamiltonian  \\
   \hline
$(\rm\expandafter{\romannumeral1})$   &      $\;\;\;\;\;\;$ $0\rightarrow\pi/2$           &$\;\;\;\;\;\;$  $0$                         &   $H^{y_1}$ \\
$(\rm\expandafter{\romannumeral2})$   &       $\;\;\;\;\;\;$ $\pi/2$                       & $\;\;\;\;\;\;$ $0\rightarrow\varphi_{c}$   &   $H^{y_2}$ \\
$(\rm\expandafter{\romannumeral3})$   &       $\;\;\;\;\;\;$ $\varphi_{c}$                 & $\;\;\;\;\;\;$ $\pi/2\rightarrow 0$        &   $H^{y_3}$ \\

\hline\hline
\end{tabular}\label{table1}
\end{table}

\subsection{Shortcuts to adiabatic single-qubit phase gate in DFS}
\label{sec42}

Here, we illustrate how to accelerate another holonomic gate, phase
gate, which is noncommuted with the single-qubit bit-phase gate. The
target Hamiltonian in the same DFS $C_{1}$ can be designed as
\begin{eqnarray}    
H^{z}_{0}(t)\;=\;\lambda'_{1,3}\,e^{i\phi}\,|a_1\rangle_{L}\langle
1|\;+\;\lambda'_{1,2}\,|a_1\rangle\langle a_2|\;+\; H.c.,
\end{eqnarray}
where $\lambda'=\sqrt{|\lambda' _{1,2}|^2+|\lambda' _{1,3}|^2}$, and
the relative phase $\theta=2\arctan(|\lambda' _{1,3}|/|\lambda'
_{1,2}|)$  and $\phi$ are the time-dependent control parameters with
$\theta\in[0,\pi]$ and $\phi\in[0,2\pi]$. The Hamiltonian
$H^{z}_{0}(t)$ has two degenerate dark states as
$|D_{0}(t)\rangle=|0\rangle_{L}$ and
$|D_{1}(t)\rangle=\cos\frac{\theta}{2}|1\rangle_{L}-\sin\frac{\theta}{2}e^{i\phi}|a_2\rangle$,
in company with two non-degenerate bright states. In the dark-state
subspace, we set $\theta=\phi=0$ initially. Using the standard
formula for the HQC, we can get the single-qubit holonomic phase
gate $U_z=e^{i\beta_{1}|1\rangle_{L}\langle 1|}$ by adiabatically
changing the angles $\theta$ and $\phi$ after a cyclic evolution,
where $\beta_{1}=-\oint\sin^{2}\frac{\theta}{2}d\phi$, corresponding
to half the solid angle swept out by the polar angles $\theta$ and
$\phi$. Thus, we can obtain the additional control Hamiltonian
$H^{z}_{1}(t)$ to realize the shortcuts to the holonomic phase gate
$U_z$ with the basis $\{|a_1\rangle,|1\rangle_{L},|a_2\rangle\}$.
That is,

\begin{eqnarray}    
H^{z}_{1}(t)&=&\frac{\dot{\phi}}{2}\sin^{2}\frac{\theta}{2}\left(
                  \begin{array}{ccc}
                    -1 & 0 & 0 \\
                    0 & 3\cos^{2}\frac{\theta}{2}-1 & -\frac{3}{2}\sin\theta e^{-i\phi} \\
                    0 & -\frac{3}{2}\sin\theta e^{i\phi} & 3\sin^{2}\frac{\theta}{2}-1 \\
                  \end{array}
                \right) \nonumber\\
             &&+\,\left(
               \begin{array}{ccc}
                 0 & 0 & 0 \\ 
                 0 & \sin^{2}\frac{\theta}{2}\dot{\phi} & \frac{1}{2}e^{-i\phi}(i\dot{\theta}+\sin\theta\dot{\phi}) \\  
                 0 & \frac{1}{2}e^{i\phi}(-i\dot{\theta}+\sin\theta\dot{\phi}) & -\sin^{2}\frac{\theta}{2}\dot{\phi} \\ 
               \end{array}
             \right).\;\;\;\;\;\;\;
\end{eqnarray}

Similar to the approach for the shortcuts to single-logic-qubit
adiabatic bit-phase gate, we can also accelerate the adiabatic
single-logic-qubit phase gate $U_z$. It can be summarized as
follows: $(\rm\expandafter{\romannumeral1})$ choosing $\phi=0$,
while changing $\theta$ from $0$ to $\pi$, the corresponding
Hamiltonian can be written by
$H^{z_1}=\lambda'\sin\frac{\theta}{2}\,|a_1\rangle_{L}\langle 1|
+\lambda'\cos\frac{\theta}{2}\,|a_1\rangle_{L}\langle a_2|+
\frac{1}{2}i\,\dot{\theta}\,|1\rangle_{L}\langle a_2|+H.c.$;
$(\rm\expandafter{\romannumeral2})$ keeping $\theta=\pi$, but
increasing $\phi$ from $0$ to $\phi_{c}$, and the required
Hamiltonian is $H^{z_2}=\lambda'\,e^{i\phi}\,|a_1\rangle_{L}\langle
1|\,+\,\lambda'\,e^{-i\phi}\,|1 \rangle_{L}\langle
a_1|\,-\,\frac{\dot{\phi}}{2}|a_1\rangle_{L}\langle
a_1|\,+\frac{\dot{\phi}}{2}|1\rangle\langle 1|$;
$(\rm\expandafter{\romannumeral3})$ setting $\phi$ unchanged, while
decreasing $\theta$ to $0$, and the required Hamiltonian takes the
form of
$H^{z_3}=\lambda'\sin\frac{\theta}{2}e^{i\phi}\,|a_1\rangle_{L}\langle
1| +\lambda'\cos\frac{\theta}{2}\,|a_1\rangle_{L}\langle a_2|+
\frac{1}{2}i\,\dot{\theta}e^{-i\phi}\,|1\rangle_{L}\langle
a_2|+H.c.$. The detailed steps for shortcuts to the phase gate are
shown in Table \ref{table2}, in which the dark eigenstates of
$H^{z}_{0}(t)$ complete a cyclic evolution in the parameter space.
Apparently, the cyclic evolution path of this approach is unlike the
one in the shortcuts to the adiabatic holonomic single-logic-qubit
bit-phase gate $U_y$. On the other hand, the quantum operations
involved for realizing the accelerated phase gate, which only
require three of the four NV centers at most to be imposed the
external classical laser pulses, are much simpler than the case in
the bit-phase gate. The experimental complexity is greatly reduced.
Based on the analysis, it is not difficult to find that as long as
the cavity frequency and the initial phase of the external laser
field are fixed, one can tune the different frequency of the
external laser pulse to achieve the shortcuts to the adiabatic
single-logic-qubit phase gate $U_z$ in DFS.

\begin{table}[htb]
\centering \caption{Scheme for a three-step approach to realize the
shortcuts to the adiabatic holonomic single-qubit phase gate.}
\begin{tabular}{cccc}
\hline\hline
  Step                                &        $\;\;\;\;\;\;$ $\theta$                     & $\;\;\;\;\;\;$ $\phi$       &   the required Hamiltonian \\
   \hline
$(\rm\expandafter{\romannumeral1})$   &      $\;\;\;\;\;\;$ $0\rightarrow\pi$           &$\;\;\;\;\;\;$  $0$                         &   $H^{z_1}$ \\
$(\rm\expandafter{\romannumeral2})$   &       $\;\;\;\;\;\;$ $\pi$                       & $\;\;\;\;\;\;$ $0\rightarrow\phi_{c}$   &   $H^{z_2}$ \\
$(\rm\expandafter{\romannumeral3})$   &       $\;\;\;\;\;\;$ $\phi_{c}$                 & $\;\;\;\;\;\;$ $\pi\rightarrow 0$        &   $H^{z_3}$ \\

\hline\hline
\end{tabular}\label{table2}
\end{table}

\section{Shortcuts to adiabatic two-qubit holonomic CP gate in DFS with NV centers system} \label{sec5}

Our shortcuts scheme for adiabatic two-qubit holonomic CP gate,
which is a more basic and crucial element for  a universal holonomic
quantum computer, is based on a variant of the proposed HQC on the
DFS in Ref. \cite{LianaoWu}. To this end, one needs eight physical
qubits to encode two logical qubits. We define four computational
states as $|00\rangle_{L}=|00010001\rangle$,
$|01\rangle_{L}=|00010010\rangle$, $|10\rangle_{L}=|00100001\rangle$
and $|11\rangle_{L}=|00100010\rangle$, with two ancillary states
$|a_3\rangle=|10000010\rangle$ and $|a_4\rangle=|01000010\rangle$.
When the physical qubits interact collectively with the dephasing
environment, the DFS  can be chosen as
$C_{2}:=span\{|00\rangle_{L},|01\rangle_{L},|10\rangle_{L},
|11\rangle_{L},|a_3\rangle,|a_4\rangle\}$, and the target
Hamiltonian takes the form as follows:
\begin{eqnarray}   \label{equation 8} 
H^{cz}_{0}(t)=\lambda'_{1,3}\,e^{i\phi}\,|a_3\rangle_{L}\langle
11|\;+\;\lambda'_{1,2}\,|a_3\rangle\langle a_4|\;+\;H.c.
\end{eqnarray}
Here the parameters $\lambda'_{1,3}$, $\lambda'_{1,2}$, and $\phi$
have the same forms as those in the case for the single-qubit
holonomic phase gate. It is straightforward to obtain the
eigenstates with zero eigenvalue of the Hamiltonian as follows:
$|D\,''_{0}(t)\rangle=|00\rangle_{L}$,
$|D\,''_{1}(t)\rangle=|01\rangle_{L}$,
$|D\,''_{2}(t)\rangle=|10\rangle_{L}$, and
$|D\,''_{3}(t)\rangle=\cos\frac{\theta}{2}|11\rangle_{L}
-\sin\frac{\theta}{2}e^{i\phi}|a_4\rangle$. The only nonzero element
of $U(4)$-valued connection is
$A_{33}=-\sin^{2}\!\frac{\theta}{2}\;\dot{\phi}$. When the dark
states evolve adiabatically along a cyclic closed path, the logical
basis $|11\rangle_{L}$ will acquire a Berry's phase $\beta_{1}$,
while the other computational components $|00\rangle_{L}$,
$|01\rangle_{L}$, and $|10\rangle_{L}$ are decoupled. The associated
two-qubit CP gate is given by
$U_{cz}=e^{\,i\,\beta_{1}\,|11\rangle_{L}\langle 11|}$ in DFS. In
our implementation, there is no need to apply four-body
interactions, just two-body ones. One can see that the adiabatic
two-qubit holonomy can be accelerated effectively, as illustrated in
the implementation for speeding up single-logical-qubit adiabatic
phase gate in DFS. Actually, this is the main advantage of our work,
different from previous works. The combination of this accelerated
two-qubit holonomic CP gate and the two noncommuting accelerated
single-qubit holonomic gates in DFS described earlier suggests that
the complete set of shortcuts to adiabatic holonomic quantum gates
in DFS are effectively built along with a realisable implementation
based on  four-NV-center systems.

Intuitively, the present scheme can be scaled up the encoded logical
qubits easily as it requires only two-body interactions. For
example, if we want to design the two-logic-qubit holonomic CP gate
between the $\emph{m}$-th and $\emph{n}$-th logical qubits, the
target Hamiltonian has the same form as Eq. (\ref{equation 8}) but
with the exchanging $\lambda'_{1,2}\rightarrow\lambda'_{4m-3,4m-2}$
and $\lambda'_{1,3}\rightarrow\lambda'_{4m-3,4m-1}$. Also, we can
realize the shortcuts for this scalable CP gate by using the
approach discussed above.

Generally, speeding up holonomic quantum gates inevitably leads to
at least an extra transition or detunings because of the existence
of the additional Hamiltonian. By taking the choice of the special
path along the geodesic curve, the controlled complexity can be
greatly reduced and the operation procedures can also be largely
simplified. For example, in the Bloch space, choosing the evolution
trajectories of the shortcuts to two noncommuting adiabatic
holonomic single-qubit gates and a two-qubit CP gate on DFS are
connected geodesic curves, one can obtain that the dynamical phases
of evolution path are vanishing, thus the set of the accelerated
adiabatic holonomic quantum gates are pure geometric. Meanwhile, in
the whole steps, a feasible route is exploited to make sure that all
of the elements of matrix-valued connection $A$ are vanishing, i.e.,
$A_{kl}^{n}=0$.

\section{Discussion and summary}\label{sec6}

For $N$ identical NV centers placed near the microsphere cavity
surface, the coupling strength between them can be expressed in
terms of the NV and cavity parameters as
$G=\Gamma_{0}|\vec{E}(r)/\vec{E}_{max}|\sqrt{V_{a}/V_{m}}$
\cite{kimble}, in which $\Gamma_{0}$ donates the spontaneous decay
rate of the excited state $|e\rangle$ for the NV center,
$|\vec{E}(r)/\vec{E}_{max}|$ is the normalized electric field
strength at the location $r$, $V_{a}=3c^3/{4\pi\nu^2\Gamma_{0}}$
serves as a characteristic interaction volume with $c$ being the
speed of light and $\nu$ being the transition frequency between the
excited state $|e\rangle$ and the ground state $|0\rangle$, and
$V_{m}$ is the cavity mode volume. The spontaneous decay rate
$\Gamma_{0}$ of the excited state reported in experiment is
$2\pi\times83~\rm MHz$ \cite{couplingstrength1, couplingstrength2}.
Considering $|\vec{E}(r)/\vec{E}_{max}|=1/6$, $\nu=471~\rm THz$ (the
transition wavelength between states $|e\rangle$ and $|0\rangle$ is
$637~\rm nm$), and $V_{m}=100~ \rm \mu m^{3}$, we obtain $G\approx
2\pi\times1~\rm GHz$ \cite{generation1}. The dephasing time of up to
$0.65~\rm ms$ for pure NV centers has been experimentally observed
\cite{dephasing}. When dynamical decoupling pulse sequences are
employed to suppress nitrogen-vacancy spin decoherence, the
dephasing time of NV centres can reach $0.6~\rm s$ at $77~\rm K$
\cite{DD}. On the other hand, the cavity frequency is
$\omega_{c}=2\pi\times74.8~\rm THz$ with the decay rate
$\kappa=2\pi\times0.0748~\rm MHz$ and the quality factor $Q=10^{9}$
\cite{Microsphere1}. The transition frequencies of NV centers are
$\omega_{10}=2.87~\rm GHz$ (zero field splitting) and
$\omega_{e0}=471~\rm THz$ with a zero-phonon line at $1.945~\rm eV$
\cite{transitionfrequency1}. Here, we have
$\omega_{10}\ll\omega_{c}$, and the detuning $\delta_{j}$ is
dependent on the difference between the cavity frequency
$\omega_{c}$ and the external laser field frequency $\omega_{L,j}$.
Choosing different frequency of classical laser field, we can obtain
the required different detuning $\delta_{j}$ when the cavity
frequency is fixed. In our implementation, the coupling strength
between the NV center and the laser field could be
$\Omega_{L}=2\pi\times500~\rm MHz$, and the detuning is
$\Delta=2\pi\times20~\rm GHz$ which satisfies the conditions
$\Delta\gg G$ and $\Delta\gg \Omega_{L}$ to ensure that the excited
state $|e\rangle$ can be eliminated adiabatically. On the other
hand, assuming $\Delta\gg \delta$, e.g., $\delta=2\pi\times2~\rm
GHz$, we have $g\approx2G\Omega_{L}/\Delta=2\pi\times50~\rm MHz$
that fulfills the large detuning condition $\delta\gg g$. This
guarantees there is no energy exchange between the NV systems and
the microcavity. Indeed, it is not necessary to apply the condition
$\Delta\gg \delta$, and we can also reach the condition of
$\delta\gg g$ provided the order of magnitude of $\Delta$ or
$\delta$ is not less than $\rm GHz$, irrespective of the relation
between them. Consequently, we can gain the different effective Rabi
frequencies $\lambda'_{j,\,k}$ by tuning the detuning between the
cavity frequency and the external laser field frequency. Once the
cavity frequency and the initial phase of the external laser field
are determined, it is easy to realize the entire physical procedures
required in the shortcuts to adiabatic HQC in DFS by changing the
external laser field frequency $\omega_{L,j}$.

\begin{figure}
\begin{center}
\includegraphics[width=8 cm,angle=0]{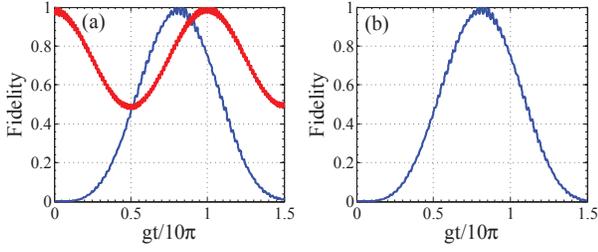}
\caption{(a) The fidelities of shortcuts to single-qubit holonomic
phase (blue line) and bit-phase (red line) gates with the initial
state $\frac{1}{\sqrt{2}}(|0\rangle_L+|1\rangle_L)$, and
$|0\rangle_L$, respectively. (b) The fidelity of shortcuts to
two-qubit holonomic CP gate with the initial state
$\frac{1}{\sqrt{2}}(|00\rangle_L+|11\rangle_L)$. Here,
$g=2\pi\times50~\rm MHz$, the decay rate of miscosphere cavity is
$\kappa=2\pi\times0.0748~\rm MHz$ \cite{Microsphere1}, and the
relaxation and dephasing rates of NV centers are
$\gamma=\gamma_\varphi=2\pi\times4~\rm KHz$ \cite{dephasing}.}
\label{fig2}
\end{center}
\end{figure}

Assuming all the qubits are in the collective dephasing environment,
we use the Lindblad master equation to simulate the performance of
the quantum gates under the influence of dissipation \cite{lindblad}
:
 \begin{eqnarray}    
\frac{d\rho}{dt}=-i[H_{int},\rho]+\kappa D[a]\rho+\gamma
D[S^{-}]\rho+\gamma_\varphi D[S^{z}]\rho,
\end{eqnarray}
where $H_{int}$ donates the Hamiltonian in the form of
Eq.~(\ref{effective}), $\rho$ is the density matrix operator,
$D[L]\rho=(2L\rho L^{+}-L^{+}L\rho-\rho L^{+}L)/2$.
$S^{-}=\sum_{i=1}\sigma_{i}^{-}$ and
$S^{z}=\sum_{i=1}\sigma_{i}^{z}$. $\gamma$ and $\gamma_\varphi$ are
the collective relaxation rate and dephasing rate of NV centers,
respectively.  $\kappa$ is the decay rate of the cavity. Here, we
define the fidelity of the gate by
$F=\langle\psi_{ideal}|\rho|\psi_{ideal}\rangle$ with
$|\psi_{ideal}\rangle$ being the corresponding ideally final state
under an ideal gate operation on its initial state
$|\psi_{in}\rangle$. Numerical simulation of the fidelities for
shortcuts to single-qubit holonomic phase, bit-phase and two-qubit
CP gates are shown in Fig.~\ref{fig2}(a) and Fig.~\ref{fig2}(b) with
initial states $\frac{1}{\sqrt{2}}(|0\rangle_L+|1\rangle_L)$,
$|0\rangle_L$ and $\frac{1}{\sqrt{2}}(|00\rangle_L+|11\rangle_L)$,
respectively. By taking the feasible experimental parameters as
$\delta_1=2\pi\times4~\rm GHz$, $\delta_2=2\pi\times0.4~\rm GHz$,
and $\delta_3=2\pi\times0.4~\rm GHz$, the fidelities of single-qubit
holonomic phase and two-qubit CP gates can reach about $99.52\%$ and
$99.76\%$, respectively. Moreover, we numerically get a high
fidelity of $99.91\%$ for single-qubit holonomic bit-phase gate with
the detunings between the frequencies of miscosphere cavity and NV
centers being $\delta_1=2\pi\times7~\rm GHz$,
$\delta_2=2\pi\times0.7~\rm GHz$, $\delta_3=2\pi\times0.7~\rm GHz$,
and $\delta_4=2\pi\times0.7~\rm GHz$. That is, our robust protocol
has a feasible physical implementation with the current experimental
techniques.


In summary, we have proposed an efficient scheme for the shortcuts
to HQC in DFS by employing TQDA. Combining the features of HQC and
DFS, the present protocol is robust against the local fluctuations
and collective noises. The optimized Hamiltonian of TQDA can greatly
shorten the time required in the adiabatic HQC to avoid the errors
due to the long runtime of quantum information processing. Moreover,
we give a feasible physical implementation of this scheme on diamond
NV centers large-detuned interacting with a quantized WGM of a
microsphere cavity. Our scheme can also be extended to
multi-logic-qubit HQC in DFS efficiently. Compared with previous
works, our scheme has the following advantages: First, the TQDA is
newly applied to implement universal HQC in DFS, and we realize
shortcuts to both two noncommuting single-qubit holonomic gates and
a two-qubit holonomic CP gate in DFS. This provides the necessary
shortcuts for the universal HQC in DFS. Second, this protocol does
not require four-body interactions and the entire quantum operation
procedures for realizing the shortcuts to universal adiabatic
holonomic quantum gates in DFS are performed by a virtual photon
process, thus the experimental challenge is much reduced. Third, our
calculation indicates that our physical implementation proposal can
be efficiently realized by appropriately applying the external laser
pulses as long as the initial conditions are determined, which
greatly simplifies the experimental complexity. Numerical
calculations reveal that the present scheme can reach a high
fidelity with current technology, which may offer a feasible route
towards robust HQC.

\bigskip
\section*{ACKNOWLEDGES}
The author Xue-Ke Song would like to thank Dr. J. Zhou for helpful
discussion. Fu-Guo Deng was supported by the National Natural
Science Foundation of China under Grant No. 11474026 and  the
Fundamental Research Funds for the Central Universities under Grant
No. 2015KJJCA01. Qing Ai was supported by
 National Natural Science Foundation of China under
Grant No. 11505007, the Youth Scholars Program of Beijing Normal
University under Grant No. 2014NT28, and the Open Research Fund
Program of the State Key Laboratory of Low-Dimensional Quantum
Physics, Tsinghua University under Grant No. KF201502.

\end{document}